\documentclass{article}
\usepackage{lscape,epsfig}
\usepackage{graphicx}
\usepackage{amssymb}
\usepackage{amsmath}
\usepackage{natbib}
\textheight=22cm \DeclareSymbolFont{ppa}{OT1}{ppl}{m}{it}
\DeclareMathSymbol{\vv}{\mathalpha}{ppa}{'166}

minus 2.3mu \thickmuskip = 2.6mu plus 2mu minus 2.6mu

\begin{document}

\newcommand{\dd}{\,{\rm d}}
\newcommand{\ie}{{\it i.e.},\,}
\newcommand{\etal}{{\it $et$ $al$.\ }}
\newcommand{\eg}{{\it e.g.},\,}
\newcommand{\cf}{{\it cf.\ }}
\newcommand{\vs}{{\it vs.\ }}
\newcommand{\zdot}{\makebox[0pt][l]{.}}
\newcommand{\up}[1]{\ifmmode^{\rm #1}\else$^{\rm #1}$\fi}
\newcommand{\dn}[1]{\ifmmode_{\rm #1}\else$_{\rm #1}$\fi}
\newcommand{\upd}{\up{d}}
\newcommand{\uph}{\up{h}}
\newcommand{\upm}{\up{m}}
\newcommand{\ups}{\up{s}}
\newcommand{\arcd}{\ifmmode^{\circ}\else$^{\circ}$\fi}
\newcommand{\arcm}{\ifmmode{'}\else$'$\fi}
\newcommand{\arcs}{\ifmmode{''}\else$''$\fi}
\newcommand{\MS}{{\rm M}\ifmmode_{\odot}\else$_{\odot}$\fi}
\newcommand{\RS}{{\rm R}\ifmmode_{\odot}\else$_{\odot}$\fi}
\newcommand{\LS}{{\rm L}\ifmmode_{\odot}\else$_{\odot}$\fi}

\newcommand{\Abstract}[2]{{\footnotesize\begin{center}ABSTRACT\end{center}
\vspace{1mm}\par#1\par \noindent {~}{\it #2}}}

\newcommand{\TabCap}[2]{\begin{center}\parbox[t]{#1}{\begin{center}
  \small {\spaceskip 2pt plus 1pt minus 1pt T a b l e}
  \refstepcounter{table}\thetable \\[2mm]
  \footnotesize #2 \end{center}}\end{center}}

\newcommand{\TableSep}[2]{\begin{table}[p]\vspace{#1}
\TabCap{#2}\end{table}}

\newcommand{\FigCap}[1]{\footnotesize\par\noindent Fig.\  %
  \refstepcounter{figure}\thefigure. #1\par}

\newcommand{\TableFont}{\footnotesize}
\newcommand{\TableFontIt}{\ttit}
\newcommand{\SetTableFont}[1]{\renewcommand{\TableFont}{#1}}
\newcommand{\MakeTable}[4]{\begin{table}[htb]\TabCap{#2}{#3}
  \begin{center} \TableFont \begin{tabular}{#1} #4
  \end{tabular}\end{center}\end{table}}

\newcommand{\MakeTableSep}[4]{\begin{table}[p]\TabCap{#2}{#3}
  \begin{center} \TableFont \begin{tabular}{#1} #4
  \end{tabular}\end{center}\end{table}}

\newenvironment{references}%
{ \footnotesize \frenchspacing
\renewcommand{\thesection}{}
\renewcommand{\in}{{\rm in }}
\renewcommand{\AA}{Astron.\ Astrophys.}
\newcommand{\AAS}{Astron.~Astrophys.~Suppl.~Ser.}
\newcommand{\ApJ}{Astrophys.\ J.}
\newcommand{\ApJS}{Astrophys.\ J.~Suppl.~Ser.}
\newcommand{\ApJL}{Astrophys.\ J.~Letters}
\newcommand{\AJ}{Astron.\ J.}
\newcommand{\IBVS}{IBVS}
\newcommand{\PASP}{P.A.S.P.}
\newcommand{\Acta}{Acta Astron.}
\newcommand{\MNRAS}{MNRAS}
\renewcommand{\and}{{\rm and }}
\section{{\rm REFERENCES}}
\sloppy \hyphenpenalty10000
\begin{list}{}{\leftmargin1cm\listparindent-1cm
\itemindent\listparindent\parsep0pt\itemsep0pt}}%
{\end{list}\vspace{2mm}}

\def\TYLDA{~}
\newlength{\DW}
\settowidth{\DW}{0}
\newcommand{\dw}{\hspace{\DW}}

\newcommand{\refitem}[5]{\item[]{#1} #2%
\def\REFARG{#3}\ifx\REFARG\TYLDA\else, {\it#3}\fi
\def\REFARG{#4}\ifx\REFARG\TYLDA\else, {\bf#4}\fi
\def\REFARG{#5}\ifx\REFARG\TYLDA\else, {#5}\fi.}

\newcommand{\Section}[1]{\section{#1}}
\newcommand{\Subsection}[1]{\subsection{#1}}
\newcommand{\Acknow}[1]{\par\vspace{5mm}{\bf Acknowledgements.} #1}
\pagestyle{myheadings}

\newfont{\bb}{ptmbi8t at 12pt}
\newcommand{\xrule}{\rule{0pt}{2.5ex}}
\newcommand{\xxrule}{\rule[-1.8ex]{0pt}{4.5ex}}
\def\thefootnote{\fnsymbol{footnote}}

\begin{center}
{\Large\bf
 The Clusters AgeS Experiment (CASE). VI.  \\
 Analysis of two detached eclipsing binaries in 
 the Globular Cluster M55\footnote{Based
 on data obtained with the Magellan, du Pont, and Swope telescopes Las Campanas Observatory.}}
 \vskip1cm
  {\large
      ~~J.~~K~a~l~u~z~n~y$^1$,
      ~~I.~B.~~T~h~o~m~p~s~o~n$^2$,
      ~~A.~~D~o~t~t~e~r$^3$
      ~~M.~~R~o~z~y~c~z~k~a$^1$,
      ~~W.~~P~y~c~h$^1$
      ~~S.~~M.~~R~u~c~i~n~s~k~i$^4$,
     and  ~~G.~~S.~~B~u~r~l~e~y$^2$
   }
  \vskip3mm
{ $^1$Nicolaus Copernicus Astronomical Center, ul. Bartycka 18, 00-716 Warsaw, Poland\\
     e-mail: (jka, mnr, wp)@camk.edu.pl\\
  $^2$The Observatories of the Carnegie Institution of Washington, 813 Santa Barbara
      Street, Pasadena, CA 91101, USA\\
     e-mail: ian@obs.carnegiescience.edu\\
  $^3$Research School of Astronomy and Astrophysics, Australian National University, 
      Canberra, Australia\\  e-mail: dotter@mso.anu.edu.au\\
  $^4$Department of Astronomy and Astrophysics, University of Toronto,
     50 St. George Street, Toronto, ON M5S 3H4, Canada\\
     e-mail: rucinski@astro.utoronto.ca}
\end{center}

\vspace*{7pt}
\Abstract
{We present an analysis of the detached eclipsing binaries V44 and V54 belonging to 
the globular cluster M55. For V54 we obtain the following absolute parameters:
$M_p=0.726\pm 0.015\,M_\odot$, $R_p=1.006\pm 0.009\,R_\odot$, $L_p=1.38\pm 
0.07\,L_\odot$ for the primary, and $M_s=0.555\pm 0.008\,M_\odot$, $R_s=0.528\pm 
0.005\,R_\odot$, $L_s=0.16\pm0.01\,L_\odot$  for the secondary. The age and apparent 
distance modulus of V54 are estimated at 13.3 -- 14.7 Gyr and $13.94\pm0.05$ mag, 
respectively. This derived age is substantially larger than ages we have derived from
the analysis of binary systems in 47 Tuc and M4.
The secondary of V44 is so weak in the optical domain that only mass function and relative 
parameters are obtained for the components of this system. However, there is a good 
chance that the velocity curve of the secondary could be derived from near-IR 
spectra. As the primary of V44 is more evolved than that of V54, such data would 
impose much tighter limits on the age and distance of M55.
}
{binaries: close – binaries: spectroscopic – globular clusters: 
individual (M55) – stars: individual (V44-M55, V54-M55)
%globular clusters: individual: M55 -- eclipsing binaries -- stars:
%individual: V44-M55, V54-M55
}

\Section{Introduction} 
The general goal of the Cluster AgeS Experiment (CASE) is to 
determine the basic stellar parameters (masses, luminosities, 
and radii) of the components of globular cluster binaries to 
a precision better than 1\% in order to measure ages and 
distances of their parent clusters, and to test stellar evolution 
models (Kaluzny et al. 2005). Within the CASE series, this is the 
third paper devoted to the study of detached eclipsing binaries
(DEBs) with main-sequence or subgiant components. In the previous 
two papers we analyzed three such systems in M4 (Kaluzny et al. 
2013) and another one in 47 Tuc (Thompson et al. 2010). 
The eclipsing binaries M55-V44 and M55-V54 (henceforth V44 and V54) 
were discovered in the field of the globular cluster M55 (Kaluzny et 
al. 2010). Follow-up spectroscopy indicated that both systems are
radial velocity members of the cluster, and their membership was 
confirmed by the proper motion study of Zloczewski et al. (2012). 

V44 and V54 are located in the turn-off region on the color-magnitude 
diagram (CMD) of the cluster. With orbital periods of 2.17~d and 
9.27~d they are well detached, and their analysis can provide 
interesting constraints on the age and distance of M55. 
In this paper we analyze light and velocity curves of the SB2 system V54 
and determine the absolute parameters of its components. We also obtain  
the mass function and a light-curve solution of the SB1 system V44. 
THe photometric and spectroscopic data are described in Section~\ref{sec:obs}.
The analysis of the data is detailed in Section~\ref{sec:analysis}. The age 
and distance of M55 are derived in Section~\ref{sec:age}, and our
results are summarized in Section~\ref{sec:sum}.\\

\section{Observations}
\label{sec:obs}
\subsection{Photometry}
\label{sec:phot}

The first eclipses of V44 and V54 were detected in 1997 during  the 
survey part of the CASE project. For the survey we used the 1-m Swope 
telescope equipped with the SITE3 $2048\times 3150$ pixel CCD camera. 
Observations of the cluster extended from 1997 to 2009. 
About 80\% 
and 20\% of the data were taken in  $V$ and $B$ bands, respectively.
At a scale of 0.435 arcsec/pixel, most of the collected images were 
adequately sampled. For frames taken with the $V$ filter the median 
seeing was 1.45 arcsec. 

Profile photometry was extracted with the 
DoPHOT package (Schechter et al. 1993), following the approach of   
Kaluzny et al. (1999). Another set of frames was collected between 
1997 and 2009 using the 2.5-m du~Pont telescope and the $2048^{2}$ pixel
TEK5 CCD camera. These observations and their reductions were described in 
detail by Kaluzny et al. (2010). Here we note that the 
time-series $BV$ photometry was performed with the image-subtraction 
method. In June 2010 the primary eclipse of V54 was additionally 
observed with the du~Pont telescope equipped with the $2048^{2}$ pixel
SITE2K CCD 
camera. All the data were calibrated onto the standard $BV$ 
system.

Altogether, several primary eclipses were observed for both V44 and V54. 
However, only for some of them were both branches (ingress and egress) 
covered, allowing for a precise timing. Secondary eclipses of 
V44 and V54 are shallow and we decided not to use them for period 
determination. Moments of primary minima  
were obtained using the KWALEX code kindly provided by Alex Schwarzenberg-Czerny. 
The O$-$C residuals listed in Table \ref{tab:min} are based on the following 
linear ephemerides:
\begin{eqnarray}
 HJD^\mathrm{V44}_\mathrm{min}& = &245 2076.86240(47) + 2.16612156(49)\times 
 E\nonumber\\
 HJD^\mathrm{V54}_\mathrm{min}& = &245 0663.6766(17) + 9.2691537(36)\times E,
 \label{eq:ephem}
\end{eqnarray}
where numbers in parentheses are the uncertainties of the last two significant 
digits. In both systems we found no evidence for any variability of the orbital 
period. Moreover, within the errors the same values for the periods were derived by 
applying the Kinmann-Lafler algorithm (Lafler \& Kinman 1965) to complete light 
curves including partially covered eclipses. Light curves obtained from 
du~Pont/TEK5 data and phased with the ephemerides (\ref{eq:ephem}) are shown 
in Fig.~1. Data obtained with the SITE2K CCD camera were used solely to find times 
of minima. 

\subsection {Spectroscopy}
\label{subsec:spectra}
Our radial velocity data are based on observations performed with
the MIKE Echelle spectrograph (Bernstein et al. 2003) on the Magellan 
Clay telescope
between July 18th, 2003 and July 12th, 2010 (UT). Most of the 
observations consisted of two 1800~s exposures interlaced with an 
exposure of a Th/Ar lamp (depending on observing conditions, some 
exposures were shorter or longer). For all observations a $0.7\times
5.0$ arcsec slit was used, and $2\times 2$ pixel binning was applied.
For $\lambda=438$~nm the resolution was $\sim$2.7 pixels at a scale 
of 0.0043~nm/pixel. The spectra were processed using a pipeline developed 
by Dan Kelson following the formalism of Kelson (2003). 

In the blue channel of MIKE (380 -- 500 nm) the typical S/N ratio  at 440 nm
was 12 for both V44 and  V54. 
The S/N ratios for the red channel (490 -- 1000 nm) spectra were markedly worse, 
and so only data from 
the blue channel were used for radial velocity measurements. For V54 
the velocities were measured using software based on the TODCOR 
algorithm of Zucker \& Mazeh (1994), kindly made available by Guillermo 
Torres.
Synthetic echelle-resolution spectra from the library of Coelho et al. 
(2005), interpolated to $\log g$ and $T_{eff}$ taken from the photometric 
solution, were used as velocity templates.  The templates  were 
Gaussian-smoothed to match the resolution of the 
observed spectra.  
We adopted ${\rm [Fe/H]} = -2.0$  
and [$\alpha$/Fe]=+0.4 for the chemical composition (Harris 1996; 
2010 electronic version). 
% Note that if you average these averages, (0.47+0.38+0.36)/3 = 0.403.
The average $\alpha$-element to Fe ratios in this cluster have been 
determined from UVES spectra of 14 red giant stars to be [Mg/Fe]=+0.47, 
[Si/Fe]=+0.38, and [Ca/Fe]=+0.36 (Carretta et al. 2009, 2010).
%
%\\{\bf Reference? JK- Reference to what? }\\
%The results of velocity measurements were insensitive to minor changes 
%in these parameters. 
%\\{\bf How minor were the minor changes?}\\
All velocities were measured on the wavelength range 400 -- 460 nm.
The results of the velocity measurements of V54 are presented in Table 
\ref{tab:rv54}. In the case of V44 even the best spectra showed only a 
marginal signal from the secondary component in the cross correlation 
function, so that velocities  have been measured for the 
primary component only. These were derived with the IRAF/FXCOR\footnote{
$^1$IRAF is distributed by the National Optical Astronomy Observatories,
which are operated by the AURA, Inc., under cooperative agreement  
with the NSF.}  task and are listed in Table \ref{tab:rv44}.  

\section {Analysis of light and radial velocity curves
%and systems parameters
}
\label{sec:analysis}

The light curves were analysed  with the PHOEBE implementation
(Pr\v{s}a et al. 2005)
of the Wilson-Devinney model (Wilson \& Devinney 1971; Wilson 1979).
Linear limb darkening coefficients were
interpolated from the tables of Claret (2000) with the help of the
JKTLD code.\footnote{Written by John Southworth and available at
www.astro.keele.ac.uk/jkt/codes/jktld.html}. The 
effective temperatures of the primary components were estimated based on 
their dereddened $B-V$ colors.  
Since the existing empirical calibrations are rather poorly 
defined for cool low-metallicity stars (see, e.g., Figs. 13 and 14 
in Casagrande et al. 2010), we employed a theoretical calibration 
based on models of Castelli \& Kurucz (2004; hereafter CK). 
For the reddening of M55 we adopted $E(B-V)=0.115\pm 0.010$ mag,
a value  0.035 mag larger than that listed in the catalog of Harris 
(1996). This choice is based on the determinations of Richter et al. 
(1999) and Dotter et al. (2010), who quote $E(B-V)=0.11\pm 0.02$ mag and 
$E(B-V)=0.115$ mag, respectively. The map of galactic extinction by 
Schlegel et al. (1998) predicts  $E(B-V)=0.136$ mag for M55; however, 
it is known to systematically overestimate the reddening (Schlafly 
et al. 2010).  Recently, a detailed reddening map for the 
region of Corona Australis complex including M55 was 
presented by Alves et al. (2014). Their map is
based on 2MASS data and has a resolution of 3 arcmin. As it turns out, the 
projected position of M55 coincides with the northeast edge
of the Pipe Nebula, see their Figs. 2 and 3. This map implies in particular 
that there is some differential reddening across the cluster field, with 
extinction decreasing from west to east.
This is in fact consistent with our $BV$ photometry based on the du Pont/TEK5 data,
which covers the central 8 arcmin of the cluster area. As one may note from 
Fig. 3, the cluster turnoff is  redder by $\approx 0.01$ mag in the region 
around V54 than around V44. 
Dr. Alves kindly communicated to us that the average reddening 
around the whole cluster amounts to $A_{V}=0.30\pm -0.01$, assuming 
$A_{V}=8.9\times A_{K}$. This in turn corresponds to 
$E(B-V)=0.097\pm 0.003$ for the standard extinction ratio of 3.1.
Given the fact V54 is located in the western part of M55 our adopted reddening
of $E(B-V)=0.11$ is consistent with results of Alves et al. (2014).
A more detailed analysis of this subject has to be postponed till the
reddening map is published. The total errors of the dereddened colors listed in Table 
\ref{tab:phot_parm} include a 0.01 mag uncertainty in the zero points 
of our $BV$ photometry.

The light curves were phased with the ephemerides 
given by eq. (\ref{eq:ephem}). For V54 we adjusted the inclination $i$, 
relative radii $r_{p}$ and $r_{s}$, effective temperature of the 
secondary $T_{s}$, 
eccentricity $e$ and periastron longitude $\omega$. In the case of V44 we 
assumed a circular orbit, so that only the first four parameters were  
adjusted. $V$ and $B$ curves were solved simultaneously. The standard 
1-$\sigma$ errors of the PHOEBE solution given in Table \ref{tab:phot_parm} 
were found using a Monte Carlo procedure written in PHOEBE-scripter and 
similar to that outlined in the description of the JKTEBOP code (Southworth 
et al. 2004, and and references therein). A total of 15,000 simulations were 
performed for each binary. Since the iterated temperature of the secondary 
was  equal to that resulting from the CK calibration to within 1$\sigma$, we 
did not attempt at any refinement of the solution. In the following, the
temperatures derived from CK calibration for for both 
components of V54  were 
used for consistency. The uncertainties given in Table \ref{tab:abs_parm} 
arise from the uncertainties in $(B-V$) listed in Table~\ref{tab:phot_parm}.

As expected from the orbital periods and the shapes of the light curves,
both systems are well detached. The largest oblateness, amounting to 0.3\%, is observed 
for the primary of V44 while for the secondary of V44 it is 
0.1\%. In the case of V54 the oblateness amounts to
0.01\% and 0.003\% for the primary and secondary, respectively. 
The residuals of the fits are shown 
in Fig. \ref{fig:VBres}, and the final values of the iterated
parameters are given in Table \ref{tab:phot_parm}. 
The derived 
relative elements are unambiguous and well defined due to the totality of 
secondary eclipses. According to the obtained solution, totality lasts 
for 0.0049~P (i.e. 1.09 h) and 0.0227 P (i.e. 1.18 h) for V44 and V54, 
respectively. 
The location of both binaries and their components on the cluster 
CMD is shown in Fig.~\ref{fig:CMD}. The primary of V44 is located right 
at the turnoff, while the primary of V54 lies 0.4-0.5 mag below the turnoff.
In both systems the secondary components are only slightly   
evolved stars, located about 4 mag below the turnoff. 

A nonlinear least-squares fit to the  observed velocity curves of V54 was 
obtained with the help of a code kindly made available by Guillermo Torres. 
While fitting, the orbital eccentricity and periastron angle were fixed based 
on the photometric solution. The observations and the orbital solutions are shown 
in Fig. \ref{fig:vc}, and the derived orbital parameters are listed in 
Table~\ref{tab:orb_parm} together with formal errors returned by the fitting 
routine. Table~\ref{tab:orb_parm} also lists the standard deviations from the 
orbital solution $\sigma_p$ and $\sigma_s$ which are a measure of the precision 
of a single velocity measurement. 
An orbital solution with eccentricity and periastron angle included among the fitted
parameters produced $e=0.1411\pm 0.0029$ and $\omega = 236.8\pm 1.0$. Both of
these values are incompatible with the light curves, but the corresponding $M\sin^3 i$
values ($M_p\sin^3 i= 0.726\pm0.014 M_\odot$ and $M_s\sin^3 i= 0.555\pm0.007) M_\odot$) 
are identical to those obtained for fixed $e$ and $\omega$ (the only difference 
is a slight reduction of the errors). Thus, the derived masses can be regarded 
as entirely reliable.   

Upon combining data 
from Tables \ref{tab:phot_parm} and \ref{tab:orb_parm} we obtained the absolute parameters 
of the components of V54 listed in Table \ref{tab:abs_parm}. 

We note that the derived systemic velocity of V54 agrees well with the radial velocity 
of M55, which according to Harris (1996) is equal to $174.7\pm 
0.3$~km/s. Lane at el. (2010) list a slightly larger value ($V_{rad}=  177.37\pm 
0.13$~km/s), but their data also show that at a projected distance of 3-5 arcmin 
from the core the velocity dispersion amounts to about 2.5 km/s. 

As V44 is a single-line binary, we could determine its mass function only.
Using data from Table \ref{tab:rv44} we obtained 
$f(m)=0.0891\pm 0.0070M_{\odot}$ and
$m_s sin(i)=0.447(m_s + m_p)^{2/3}\pm 0.017 M_{\odot}$
with the rms of residuls from the fit amounting to 1.9 km/s.
The derived systemic velocity, $V\gamma=179.05\pm 0.46$ km/s, confirms
the radial-velocity membership of V44 in M55 (Lane at el. 2010).

In the $V$ band the primary of V44 is by 0.5 mag brighter the primary of V54.
Based on Dartmouth isochrones (Dotter et al. 2008), this implies a difference 
in masses of about 0.02~$M_{\odot}$ at the turnoff of M55.
Hence, the mass of the primary of V44 can be estimated at $m_{p}^{V44}\approx 
0.74 M_{\odot}$. From the mass function one gets $m_{s}^{V44}\approx 0.53 
M_{\odot}$ - a value consistent with $m_{s}^{V54}$ and the location of the 
secondaries of V44 and V54 on the CMD of the cluster (Fig. \ref{fig:CMD}).  

\Section{Age and distance of V54}
 \label{sec:age}

% DONE Age from mass-radius and mass-luminosity relations.\\ 
%Note the strong dependence of Lbol on the adopted calibration Teff/B-V.
% DONE Emphasize the agreement of the isochrone fit to CMD.\\
%Which calibration do we use? Possibly an average of all 
%with the dispersion of Teff treated as additionall uncertinty.\\
%Can we exclude  Y=0.274 claimed by Alvarez and Sandquist (2007)?\\
%Distance from Mv once we agree on Teff.

The fundamental parameters listed in Table \ref{tab:abs_parm} allow for a
determination of the age of M55 via comparison with stellar isochrones. The
following analysis has been carried out using isochrones from Dartmouth 
(Dotter et al. 2008) and Victoria-Regina (VandenBerg et al. 2012). These 
isochrones share similar physical assumptions, including nuclear reaction
rates and opacities. The most significant difference is that the Dartmouth
models include the effects of the diffusion of He and metals whereas the Victoria-Regina 
models include the diffusion of He but not metals. In both cases, the full effect 
of diffusion is mitigated near the surface (see the respective papers for further 
details). The Dartmouth isochrones adopt a constant $\alpha$-enhancement
factor (in this case [$\alpha$/Fe]=+0.4 is assumed) whereas the Victoria-Regina 
models are based on an observationally-motivated, element-specific $\alpha$-enhancement, 
amounting to [$\alpha$/Fe] $\simeq +0.33$. 

As the results obtained from the two sets of models turned out to be consistent with 
each other at $1\sigma$-level, in the following we only report those based on Dartmouth 
isochrones. Fig. \ref{fig:dartmouth} shows a side-by-side comparison of the mass-radius 
plane containing V54 and the CMD from a region surrounding V54, to which we fitted isochrones 
using the reddening value $E(B-V)=0.115$ from Section~\ref{sec:analysis}. Qualitatively,
the $M-R$ diagram indicates an age of 14-15 Gyr for the primary and a younger age for the 
secondary. The CMD comparison favors an age of 13-14 Gyr - in agreement with the mass-radius 
relation of the secondary. We note that the age derived from the mass-radius relation is 
free from uncertainties in distance and extinction that plague ages derived via isochrone 
fitting to the CMD. 

A quantitative age analysis was performed for the components V54 following the 
technique described by Kaluzny et al.\ (2013) to study three binary systems 
in M4. Briefly, the method densely samples the $(M-R)$ or $(M-L)$ plane  
in both dimensions for each component separately, and finds the age of an isochrone 
that matches the properties of the sampled point. Thus generated age ensembles may 
be visualized in the form of histograms, as in Kaluzny et al.\ (2013), or mass-age 
diagrams (Fig. \ref{fig:ellipses}). The latter show ellipses composed of points which 
on $(M-R)$ or $(M-L)$ plane are $1\sigma$-distant from the points specified by 
the solutions given in Table \ref{tab:abs_parm}, where $\sigma^2 = \mathrm{err}(M/M_\odot)^2
+\mathrm{err}(R/R_\odot)^2$ for the $(M-R)$ plane and likewise for the $(M-L)$ plane; 
$\mathrm{err}(M/M_\odot)$, $\mathrm{err}(R/R_\odot)$ and $\mathrm{err}(L/L_\odot)$ being 
the standard errors of $M$, $R$ and $L$ from Table \ref{tab:abs_parm}. 

The ellipses for the primary in Fig. \ref{fig:ellipses} are much smaller than those for
the secondary, setting tighter limits for the age of V54. This is to be expected, as the 
primary sits just below the main sequence turnoff and evolves relatively quickly, while the 
secondary is located well down the `unevolved' main sequence (Fig.~\ref{fig:CMD}), where 
the structure of the star changes very slowly with time. Each of the four ellipses in Fig. 
\ref{fig:ellipses} defines a range of ages achieved by stars whose parameters differ from 
those listed in Table \ref{tab:abs_parm} by at most one standard error. Thus, a fair 
estimate of the age of V45 is the common part of the four ranges. The limits 13.3 -- 14.7 
Gyr derived in this way are compatible with the age of 13.5$\pm$1.0 Gyr obtained by Dotter et al. (2010)
by isochrone fitting to HST data.
 
For stellar parameters from Table \ref{tab:abs_parm}, the CK calibration gives bolometric 
corrections in $V$ of $0.126\pm0.005$ mag for the primary and $0.230\pm0.020$~mag 
for the secondary. Using $M_\odot^{bol} = 4.755$\footnote{
Mamajek's Star Notes, 
https://sites.google.com/site/mamajeksstarnotes/basic-astronomical-data-for-the-sun}
, we get absolute $V-$luminosities  
$M_{Vp} = 4.53\pm0.05$ mag and $M_{Vs} = 6.93\pm0.08$ mag, corresponding to apparent 
distance moduli $\mu_p = 13.87\pm0.05$ mag and $\mu_s = 14.00\pm0.08$ mag. The mean 
modulus of 13.94$\pm$0.05 mag agrees reasonably well with 14.03 mag obtained from CMD 
fitting and with 13.88 mag found by Dotter et al. (2010). Assuming $A_V = 3.1\times 
E(B-V)$, the dereddened modulus is 13.58 mag. 

The combination of CMD and mass-radius relation provide some leverage on the He content
of M55. An initial He mass fraction of Y$\sim0.25$ has been assumed in the preceding
analysis. However, there is at least one claim of a higher initial He mass fraction, 
Y=$0.274\pm0.016$, based on star counts and the R-parameter (Vargas Alvarez \& Sandquist 2007).
In order to test this claim, the analysis carried out above was redone using Dartmouth 
isochrones with Y=0.27. Ages derived for Y=0.27 from the mass-radius relation yield a 1.8 Gyr 
reduction in the age of the primary. This is entirely consistent with composition-sensitivity
tests performed by Kaluzny et al.\ (2013; their Section 6 and Table 14) and we refer the reader to 
that paper for further discussion of changes in chemical composition to the ages determined
via mass-radius analysis. 
The position of isochrones through the turnoff and subgiant branch in the CMD are 
insensitive to modest changes in the initial He content at fixed age and metallicity
(see, e.g., Figure 1 of Dotter, Kaluzny, \& Thompson 2009). Hence the CMD of M55 will still
favor an age of $\sim14$ Gyr with Y=0.27. The consistency found between mass-radius, mass-luminosity, 
and CMD fitting described in this section argues in favor of an initial Y$\sim0.25$ for M55.

\Section{Summary}
 \label{sec:sum}

We have analysed photometric and spectroscopic observations of the  eclipsing binaries 
V44 and V54 in M55. V54 is an SB2 system, and we obtained absolute parameters of the
components of this system, leading to an estimate of age and distance modulus of the 
cluster. The resulting age of 13.3 -- 14.7 Gyr derived from $M-R$ and $M-L$ diagrams is compatible with 
the age 13 -- 14 Gyr obtained from CMD fitting with Dartmouth or Victoria-Regina isochrones. 
Apparent distance moduli derived with the same two methods and the same two sets of isochrones 
are also compatible within the errors.
 
The best chance to tighten the age limits is offered by the mass-radius relation, which is 
independent of uncertainties in distance and extinction that plague ages obtained via isochrone 
fitting to the CMD. Since at the location of the primary on the $M-R$ plane the isochrones are 
almost perpendicular to the mass axis (that's why the corresponding ellipses in Fig. 
\ref{fig:ellipses} are very narrow), the age uncertainty arises mainly from the error in the 
mass of the primary. The only way to reduce it is by taking additional spectra of V54. Thompson 
et al. (2010) estimate that doubling of the present set of velocity measurements would improve 
the mass estimates by 33 per cent.

The uncertainty of the location of V54 components on the $M-L$ plane is mainly due to the 
relatively poor accuracy of their luminosities which were found using absolute radii from 
Table \ref{tab:abs_parm} and effective temperatures estimated from $(B-V)-T_{eff}$ calibration 
of Castelli and Kurucz (2004). Unfortunately, the existing empirical calibrations are poorly 
defined for low mass stars  with ${\rm [Fe/H]<-1.5}$ (Casagrande et al. 2010; Gonzalez-Hernandez 
\& Bonifacio 2009). A significant improvement is expected when results from the  GAIA astrometric 
mission will become publicly available. Also our determination of the distance to M55 depends heavily
on the adopted teperature scale. A better accuracy could be achieved based on the surface brighness 
in $V$, which is known to tightly correlate with the $V-K$ color (Di Benedetto 2005). Given that V54 
is located in the outer uncrowded part of the cluster, accurate near-IR photometry 
of these systems   
with large telescopes is entirely feasible.

Independent and potentially more accurate estimates of age and distance of M55 may follow  
from the analysis of the V44 system, whose primary is more evolved than the primary of V54 (see
Fig.~\ref{fig:CMD}). While the secondary of V44 is too faint to measure its velocity in the 
visible domain, it seems to contribute enough light for successful measurements in the near 
IR. Accurate parameters of both systems could yield not only tight age limits, but also 
impose useful constraints on the primordial He content of the cluster.

The age we derived for M55 is by $\sim2$ Gyr older than the ages we
obtained from the analysis of eclipsing binaries in  47~Tuc (Thompson et
al. 2010) and M4 (Kaluzny et al. 2013). This can be compared with recent
result of Hansen et al. (2013) who, based on cooling sequences of white
dwarfs, found that NGC 6397 ([Fe/H]=-2) is about 2 Gyr older than 47 Tuc.
 Hansen et al. stress the benefits of deriving ages from WDs, which are
largely insensitive to metallicity, but are still subject to uncertainties
in distance and reddening.  In our case, sensitivity to metallicity is
present but the uncertainties in distance and reddening are not.  Since M55
and NGC 6397 have very similar metallicities, it is significant that these two
more-or-less independent methods arrive at a similar age difference between
metal-poor and metal-rich globular clusters.

\Acknow
{
JK, MR and WP were partly supported by the grant DEC-2012/05/B/ST9/03931
from the Polish National Science Center. We thank Dr. Joao Alves for
sending us the information about reddening in the M55 region.
This series of papers is dedicated 
to the memory of Bohdan Paczy\'nski.
}

\clearpage

\begin{table}
 \caption{Times of minima for V44 and V54. \label{tab:min}}
 \centering
 \begin{tabular}{@{}lrrrr}
 \hline
 \hline
Star & E & Tmin [d]    & $\sigma$ [d] & O-C [d] \\
     &   & {\footnotesize HJD-2450000} &              &         \\ 
\hline
V44&  -687	& 588.73725&      0.00137	& -0.00036\\
V44&  -674	& 616.89698&	  0.00130	& -0.00051\\
V44&	0     &  2076.86429& 	  0.00173	& -0.00189\\
V44&   319 &     2767.85376& 	  0.00140	&  0.00142\\
V44&   366 &     2869.66077& 	  0.00101	&  0.00212\\
V44&   836 &     3887.74008& 	  0.00070	& -0.00006\\
V44&   849 &     3915.89977& 	  0.00039	& -0.00016\\
V44&   871 &     3963.55348& 	  0.00390	&  0.00080\\
V44&   872 &     3965.72231& 	  0.00138	& -0.00191\\
V44&  1023 &     4292.80459& 	  0.00066	&  0.00016\\
V44&  1179 &     4630.71970& 	  0.00037	&  0.00002\\
   &       &               &                    &         \\
V54&	 0 &	  663.67449& 	  0.00213	&  0.00215\\
V54&   238 &     2869.73600&      0.00173       & -0.00078\\
V54&   428 &     4630.87478&      0.00062       & -0.00035\\
V54&   428 &     4630.87550&      0.00112       & -0.00107\\
V54&   465 &     4973.83301&      0.00048       &  0.00011\\
V54&   465 &     4973.83449&      0.00104       & -0.00137\\
V54&   506 &     5353.86830&      0.00052       &  0.00012\\
V54&   506 &     5353.86750&      0.00069       &  0.00092\\
\hline
\end{tabular}\\
\end{table}

\clearpage

\begin{table}
\caption{Radial velocity observations of V54 
\label{tab:rv54} 
     }
 \centering
 \begin{tabular}{@{}lrrrrr}
 \hline
 \hline
HJD      & $v_{p}$   &  $v_{s}$ & Phase  & $(O-C)_{p}$ & $(O-C)_{s}$\\
-2450000 & [km/s]    &  [km/s]  &        & [km/s]      & [km/s] \\
%\\
\hline
2871.6532 &125.36 &237.37 &0.207 &  -0.86&  -0.53\\
2872.6607 &126.90 &238.86 &0.316 &  -0.74&   2.82\\
2922.5065 &219.00 &118.64 &0.693 &  -0.26&   2.37\\
2923.4957 &210.74 &127.98 &0.800 &  -0.51&   1.24\\
2944.5042 &154.07 &202.33 &0.066 &  -0.58&   1.59\\
2946.4997 &121.87 &244.55 &0.282 &  -1.90&   3.44\\
3178.7800 &132.97 &226.52 &0.341 &   0.18&  -2.79\\
3180.7962 &205.80 &135.77 &0.559 &   0.86&   0.78\\
3182.8380 &213.75 &121.85 &0.779 &  -0.07&  -1.53\\
3206.6655 &134.53 &224.35 &0.350 &  -0.35&  -2.23\\
3210.6248 &212.69 &126.03 &0.777 &  -1.38&   2.97\\
3280.4991 &127.70 &236.88 &0.315 &   0.13&   0.74\\
3281.5346 &161.52 &189.97 &0.427 &   0.76&  -2.77\\
3282.5326 &198.61 &144.62 &0.535 &  -0.10&   1.49\\
3581.7421 &209.19 &134.09 &0.815 &  -0.01&   4.67\\
3582.7236 &189.51 &156.47 &0.921 &  -0.17&   1.53\\
3584.6869 &139.41 &222.42 &0.132 &   0.26&   1.42\\
3585.6744 &124.05 &245.53 &0.239 &   0.54&   4.09\\
3632.5485 &124.88 &241.14 &0.296 &  -0.10&   1.61\\
3635.5729 &215.62 &122.48 &0.622 &  -0.19&   1.69\\
3875.8479 &203.24 &138.00 &0.544 &   1.91&  -1.71\\
3877.8737 &215.39 &120.77 &0.763 &  -0.12&  -0.40\\
3891.8346 &124.06 &243.21 &0.269 &   0.89&   1.33\\
3937.6913 &125.74 &242.07 &0.216 &   0.52&   2.86\\
4648.7677 &186.33 &162.43 &0.930 &  -1.18&   4.66\\
4656.8090 &211.28 &127.79 &0.798 &  -0.23&   1.39\\
5389.7060 &199.92 &143.79 &0.866 &  -0.70&   3.15\\
\hline
\end{tabular}
\end{table}

\begin{table}
\caption{Radial velocity observations of V44
\label{tab:rv44}
     }
 \centering
 \begin{tabular}{@{}lrlr}
 \hline
 \hline
HJD      & $v_{p}$   &  HJD       & $v_{p}$ \\
-2450000 & [km/s]    &  -2450000  &  [km/s]  \\
%\\
\hline
2838.6719 & 253.40 & 3892.8097 & 118.47\\
2839.6845 & 112.64 & 3937.7585 & 138.68\\
3179.9013 & 105.55 & 3937.8037 & 130.17\\
3180.8638 & 241.95 & 4258.8596 & 113.87\\
3280.5677 & 251.39 & 4314.7773 & 123.18\\
3281.6041 & 113.14 & 4333.5189 & 249.88\\
3282.5920 & 231.30 & 4655.7762 & 204.85\\
3580.7507 & 106.60 & 4656.7739 & 168.57\\
3877.8054 & 141.95 & 5038.7444 & 123.71\\
3891.7597 & 237.46 &           &       \\
\hline
\end{tabular}
\end{table}

\begin{table}
 \caption{Photometric parameters of 54 and V44$^\mathrm{a}$
          \label{tab:phot_parm}
         }
\centering
 \begin{tabular}{lcc}
  \hline
   Parameter        & V54 & V44    \\
  \hline
     $i$(deg)       &   89.050(27)&  86.86(11)  \\
     $r_p$                &0.04992(32)  & 0.15835(54)   \\
     $r_s$                & 0.02617(23) & 0.06807(45)   \\
     $e$            &  0.133(5) &  0$^\mathrm{b}$ \\
     $\omega$ (deg) &  237.1(2)   &  0$^\mathrm{b}$  \\
     $T_{p}$ (K)               & 6246$^\mathrm{b}$ &  6602$^\mathrm{b}$ \\
     $T_{s}$ (K)$^\mathrm{c}$  & 4959(16) &  4807(28) \\
     $(L_{p}/L_{s})_{V}$  & 10.80(23)  & 0.2326(54)     \\
     $(L_{p}/L_{s})_{B}$  & 14.93(39)  & 0.     \\
     $\sigma_{rms}(V)$ (mmag) & 83  & 75    \\
     $\sigma_{rms}(B)$ (mmag) & 101  & 75   \\
     $V_p$ (mag)$^\mathrm{d}$ & 18.399(7)  & 17.928(7) \\
     $V_s$ (mag)$^\mathrm{d}$ & 20.983(22) & 21.391(27) \\ 
     $B_p$ (mag)$^\mathrm{d}$ & 18.933(6)  & 18.417(5) \\
     $B_s$ (mag)$^\mathrm{d}$ & 21.869(26) & 22.370(34) \\
     $(B-V)_{0p}$(mag)$^\mathrm{e}$& 0.419(14) & 0.374(18) \\ 
     $(B-V)_{0s}$(mag)$^\mathrm{e}$& 0.771(32) & 0.864(45) \\ 
  \hline
 \end{tabular}\\
\rule{0 mm}{3 mm}
$^\mathrm{a}$Numbers in parentheses are the errors of the 
last significant digits.\\
\rule{1 mm}{0 mm}$^\mathrm{b}$Assumed in fit.\\
\rule{1 mm}{0 mm}$^\mathrm{c}$Result of PHOEBE iterations.\\
\rule{1 mm}{0 mm}$^\mathrm{d}$The internal error from the photometric solution and 
profile photometry is given.\\
\rule{1 mm}{0 mm}$^\mathrm{e}$The error includes the internal error from the photometric 
solution and profile photometry as well as errors of zero point of photometry
and interstellar extinction. 
\end{table}

\begin{table}
 \caption{Orbital parameters of V54$^\mathrm{a}$
          \label{tab:orb_parm}
         }
 \centering
 \begin{tabular}{lc}
  \hline
   Parameter        & Value    \\
  \hline
  $\gamma$ (km s$^{-1}$)     & 174.62(15)  \\
     $K_p$ (km s$^{-1}$)     & 48.13(19)  \\
     $K_s$ (km s$^{-1}$)     & 62.92(60)  \\
     $e$                       & 0.133$^{\mathrm b}$ \\
     $\omega$ (deg)            & 237.1$^{\mathrm b}$ \\
     $\sigma_p$ (km s$^{-1}$)   & 0.79             \\
     $\sigma_s$ (km s$^{-1}$)   & 2.50              \\
     Derived quantities:        &                 \\
     $A\sin i$ (R$_\odot$)     & 20.16(12)  \\
     $M_p\sin^3 i$ (M$_\odot$) & 0.7259(154)   \\
     $M_s\sin^3 i$ (M$_\odot$) & 0.5552(75)  \\
\hline
 \end{tabular}\\
\rule{0 mm}{3 mm}
$^\mathrm{a}$Numbers in parentheses are the errors of the
last significant digits\\
\rule{1 mm}{0 mm}$^\mathrm{b}$Assumed in fit\\
\end{table}

\begin{table}
 \caption{Absolute parameters of V54
          \label{tab:abs_parm}
         }
 \centering
 \begin{tabular}{ll}
  \hline
    Parameter        & Value    \\
  \hline
     $M_p$ (M$_\odot$) & 0.726(15)   \\
     $M_s$ (M$_\odot$) & 0.555(8)  \\
     $R_p$ (R$_\odot$) & 1.006(9)   \\
     $R_s$ (R$_\odot$) & 0.528(5)  \\
     $T_p$ (K)$^a$ & 6246(71)   \\
     $T_s$ (K)$^a$ & 5020(95)  \\
     $L^{bol}_{p} $ (L$_\odot$) & 1.381(67)  \\
     $L^{bol}_{s} $ (L$_\odot$) & 0.159(12)  \\
\hline
 \end{tabular}\\
\rule{0 mm}{3 mm}
$^a$ Derived from the calibration of Castelli \& Kurucz (2004). 
The errors reflect the uncertainties in $(B-V)$ from 
Table \ref{tab:phot_parm}.
\end{table}

\clearpage

\begin{figure}
 \centerline{\includegraphics[width=0.95\textwidth,
             bb= 25 175 442 718,clip ]{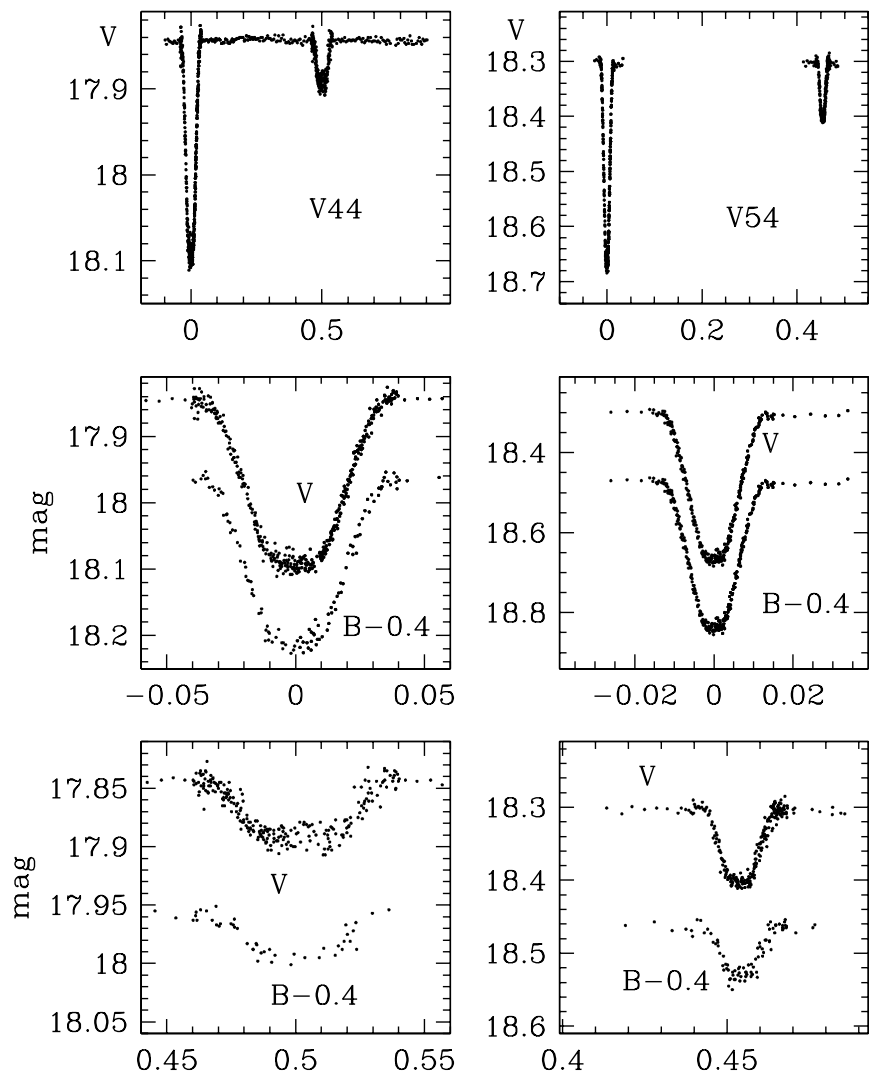}}
 \caption
  {Light curves of V44 (left column) and V54 (right column) obtained with 
the du~Pont telescope and the TEK5 CCD camera. The $B$-curves
   are shifted upwards by 0.4 mag.
   \label{fig:lc}
  }
\end{figure}

\clearpage

\begin{figure}
 \centerline{\includegraphics[width=0.95\textwidth,
             bb= 29 419 564 690,clip ]{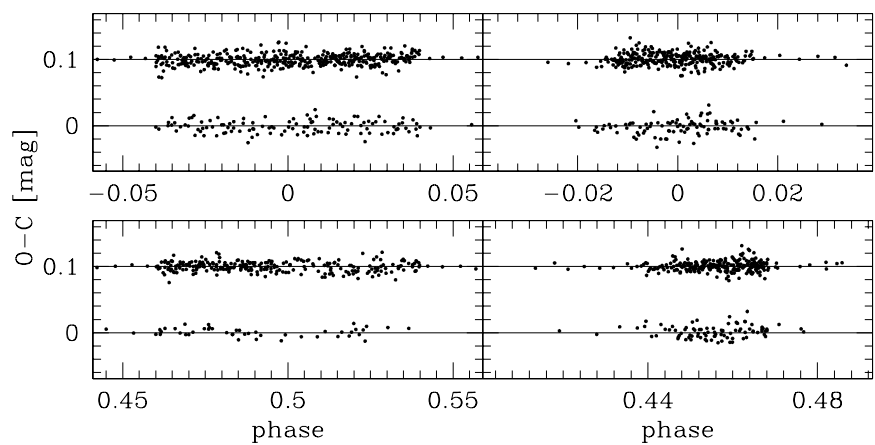}}
 \caption
  {
Residuals to the fits adopted for the analysis. Left panels: V44. 
Right panels: V54.
 In each panel, the lower sequence represents the $B$-residuals, 
and the upper one the $V$-residuals offset by 0.1 mag for clarity.
   \label{fig:VBres}
  }
\end{figure}

\begin{figure}
 \centerline{\includegraphics[width=0.95\textwidth,
             bb= 22 16 596 784,angle=270,clip ]{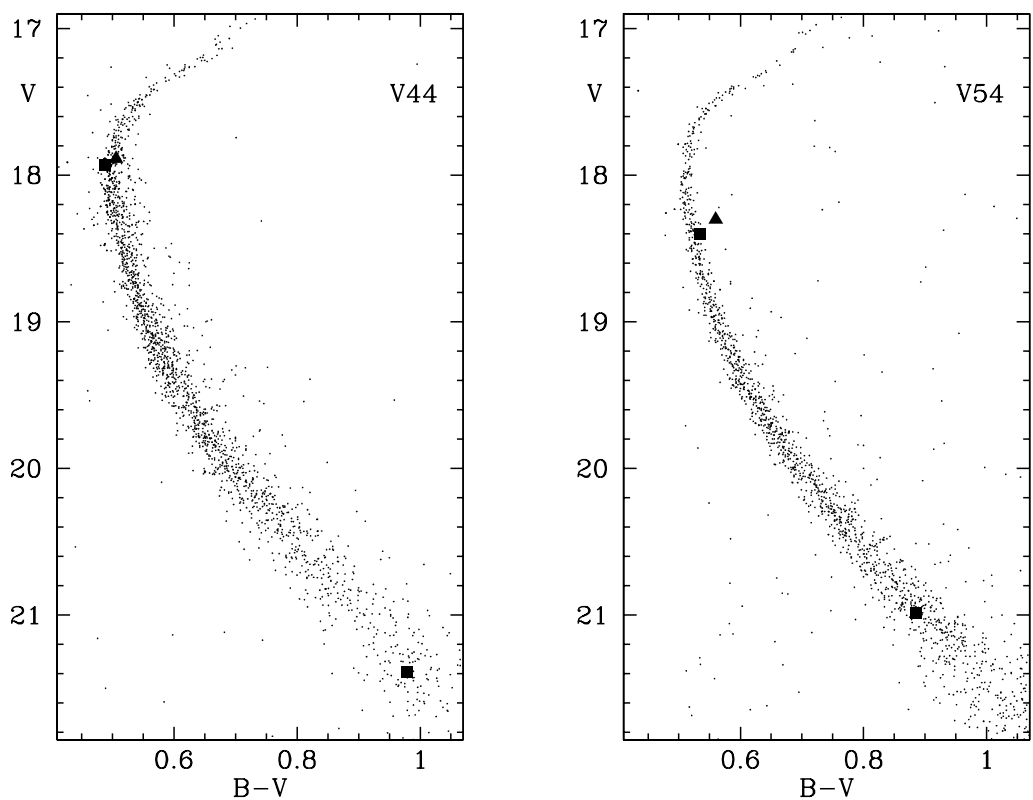}}
 \caption
  {Color-magnitude diagrams of M55 with marked positions 
of V44 and V54 systems (triangles) and their components (squares). 
Stars in each panel are taken from the surroundings of the 
corresponding system. 
   \label{fig:CMD}
  }
\end{figure}

\clearpage

\begin{figure}
 \centerline{\includegraphics[width=0.95\textwidth,
             bb= 29 176 564 515,clip ]{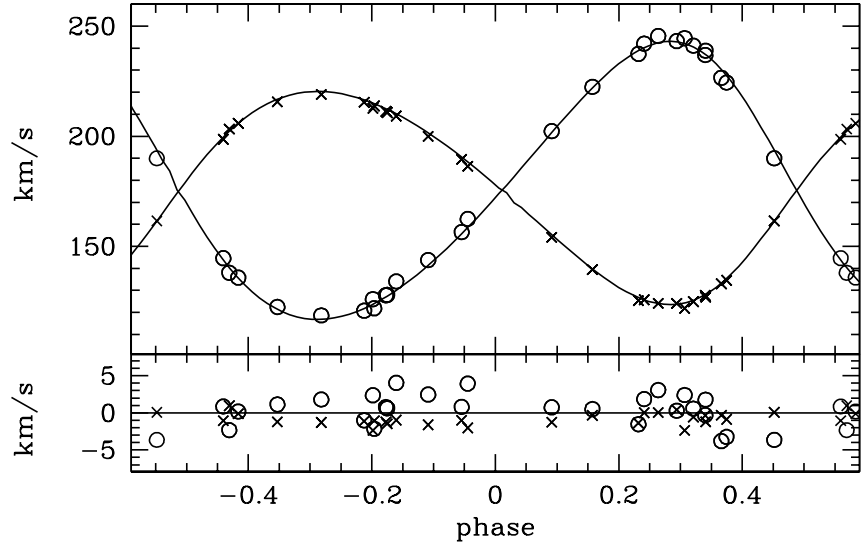}}
 \caption
  {Velocity curve of V54 adopted for the analysis. Crosses and open circles denote, 
   respectively, primary and secondary velocities (upper panel) and residuals to 
   the fit (lower panel).
   \label{fig:vc}
  }
\end{figure}

\begin{figure}
   \centerline{\includegraphics[width=0.95\textwidth,
               bb = 32 323 565 690]{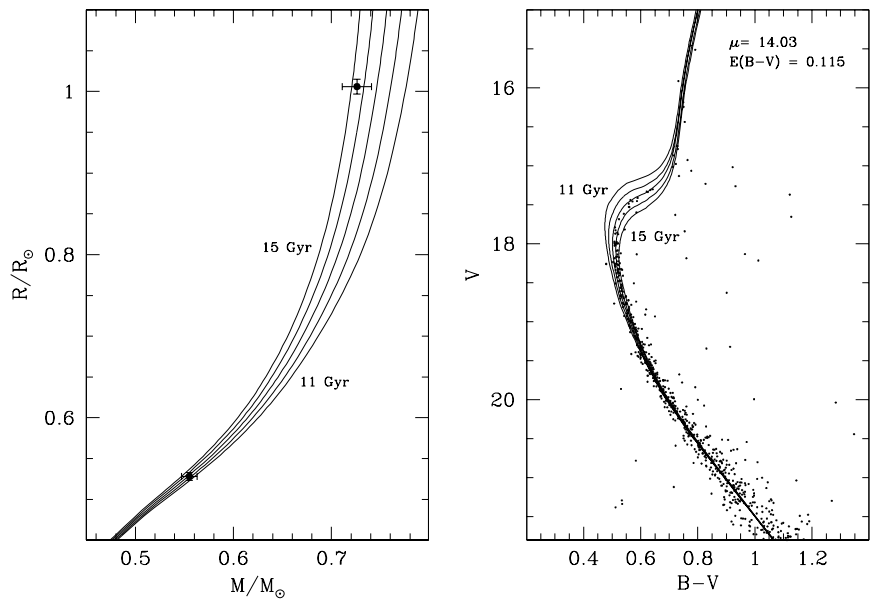}}
   \caption{Left: Dartmouth isochrones (lines) at [Fe/H]=$-2.0$, 
    [$\alpha$/Fe]=+0.4 compared to the components of V54 (points 
    with $1\sigma$ errors) in the mass-radius diagram. 
    Right: The same isochrones transformed to the observational 
    plane compared to the CMD from a region of M55 containing V54. 
    The apparent distance modulus resulting from the fit  is equal 
    to 14.03 mag for 
    $E(B-V)$=0.115 mag. To better visualize 
    the isochrones, the stars shown are from a smaller area surrounding of 
    V54 than the region used for the right panel of Fig. 3.
    \label{fig:dartmouth}}
\end{figure}

\clearpage

\begin{figure}
 \centerline{\includegraphics[width=0.95\textwidth,
%             bb= 28 400 564 690,clip ]{mrl.eps}}
             bb= 28 320 564 690,clip ]{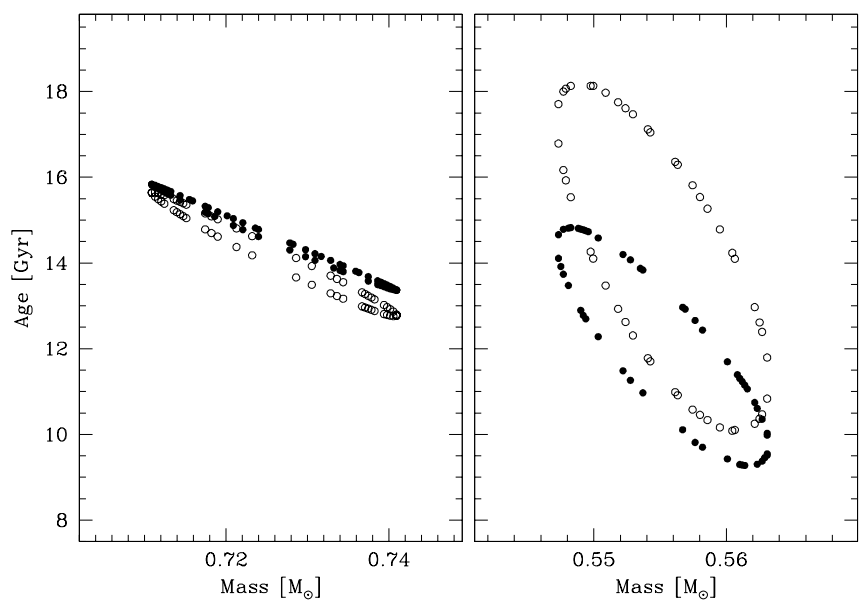}}
 \caption{Mass-age diagrams for the primary (left) and the secondary (right) component 
   of V54 based on Dartmouth isochrones. Results from the analysis of $(M-R)$ and $(M-L)$ 
   planes are shown with dots and open circles, respectively (see Section~\ref{sec:age} for
   a detailed explanation).
   \label{fig:ellipses}}
\end{figure}

\end{document}